\documentclass[floatfix,pra,twocolumn,aps,superscriptaddress,showpacs]{revtex4-2}
\usepackage{graphicx}
\usepackage[normalem]{ulem}
\usepackage{mathrsfs,amsmath}
\usepackage{amsfonts}
\usepackage{amssymb}
\usepackage{amsmath}
\usepackage{epsfig}
\usepackage{subfigure}
\usepackage{mathtools}
\usepackage{braket}
\usepackage[usenames,dvipsnames]{color}
\usepackage{setspace}
\usepackage{bm}
\usepackage{times}
\usepackage{ulem}
\usepackage{dsfont}
\usepackage{xcolor}
\usepackage{adjustbox}
\usepackage{hyperref}
\usepackage{soul}
\hypersetup{
	colorlinks=true,
	citecolor=blue,
	linkcolor=blue,
	urlcolor=blue}

\begin{document}
\title[]{Metastability, collective excitations, and traveling states of one-dimensional spin-orbit-coupled quantum droplets}

\author{Harsimranjit Kaur}
\thanks{These authors contributed equally to this work.}
\affiliation{Department of Physics, Central University of Rajasthan, Ajmer 305817, India}
\author{Ritu}
\thanks{These authors contributed equally to this work.}
\affiliation{Department of Physics, Indian Institute of Technology Ropar, Rupnagar 140001, Punjab, India}
\author{Rajat}
\email{rajat13sep@gmail.com}
\affiliation{Department of Physics, Indian Institute of Technology Ropar, Rupnagar 140001, Punjab, India}
\affiliation{Department of Physics, Ben-Gurion University of the Negev, Beer-Sheva 84105, Israel} 
\author{Kuldeep Suthar}
\email{kuldeep.suthar@curaj.ac.in}
\affiliation{Department of Physics, Central University of Rajasthan, Ajmer 305817, India}
\author{Sandeep Gautam}
\email{sandeep@iitrpr.ac.in}
\affiliation{Department of Physics, Indian Institute of Technology Ropar, Rupnagar 140001, Punjab, India}

\begin{abstract}

We study the static and traveling droplet phases of spin-orbit coupled, one-dimensional Bose-Bose mixtures with a Lee--Huang--Yang correction. For the static droplets, 
employing a variational analysis for an infinitely extended mixture, we show that the system can feature two ground-state droplet phases: a stripe droplet phase up to a 
critical Rabi coupling strength, at which it undergoes a first-order transition to a zero-momentum droplet phase. The variational results for the ground-state properties are shown to 
be in good agreement with numerical solutions for droplets in the thermodynamic limit as well as finite-size droplets in the ``flat-top regime''. For finite-size droplets, the 
stripe phase can feature multiple stationary states with different numbers of peaks in their stripe density patterns. The 
calculation of the excitation spectrum for these stationary states confirms their stability. To study the moving droplets, we investigate the ground-state phases of the system 
in the droplet's comoving frame. We show that in the comoving frame, droplets acquire a finite longitudinal magnetization. As a consequence, the phase diagram features two 
distinct phases: a traveling stripe droplet and a plane-wave droplet. The real-time evolution of these droplets in the space-fixed frame, complemented by the Bogoliubov analysis 
in the comoving frame, illustrates the stability of these traveling
droplet phases. 

\end{abstract}
\maketitle


\section{Introduction}\label{Sec.I}
Quantum droplets are self-bound dilute liquids in which beyond-mean-field quantum fluctuations arrest a mean-field instability. The mechanism was first formulated for attractive 
Bose-Bose mixtures by Petrov~\cite{petrov_15,petrov_notes}, and was subsequently realized in homonuclear and heteronuclear mixtures~\cite{cabrera_18, semeghini_18,cheiney_18, 
PhysRevLett.122.090401, derrico_19,Wang_PhysRevResearch.3.033247, Burchianti_PhysRevLett.134.093401}, complementing the earlier observation in dipolar quantum 
droplets~\cite{Kadau2016,PhysRevLett.116.215301,schmitt_16,chomaz_16}. Dimensionality alters the balance that governs self-binding. In one dimension, the leading Bogoliubov 
correction is attractive; together with a weak mean-field repulsion, it produces a finite-density global minimum within the mean-field stability 
regime~\cite{petrov_16,10.21468/SciPostPhys.19.5.133}. Quantum Monte Carlo calculations have tested the validity of the weak-coupling description~\cite{PhysRevLett.122.105302, 
PhysRevA.102.023318}, while studies of finite systems have revealed distinct density and spin excitations~\cite {PhysRevA.102.023318, PhysRevA.98.013631, PhysRevA.101.051601, 
w68w-68w5}. The bulk of these conventional Bose-Bose droplets is spatially uniform without any density modulations.

Raman-induced spin-orbit (SO) coupling introduces a tunable momentum scale by locking pseudospin to momentum. Following its experimental realization~\cite{Lin2011}, interacting SO-coupled Bose gases were shown to support stripe (ST), plane-wave (PW), and zero-momentum (ZM) phases~\cite{PhysRevLett.105.160403, PhysRevLett.107.150403, PhysRevLett.108.225301}. Stripe order has been measured directly~\cite{li2017stripe, PhysRevLett.124.053605}, and its supersolid excitation spectrum has been studied theoretically~\cite{PhysRevLett.110.235302, *PhysRevA.95.033616, *PhysRevLett.127.115301, *PhysRevA.111.033316} and recently measured experimentally~\cite{chisholm2024probing}.

The combination of spin-orbit (SO) coupling and quantum fluctuation-induced self-binding has motivated several studies of SO-coupled quantum liquids. A first-order ST-to-ZM transition was predicted for a two-dimensional (2D) self-bound gas as the Raman coupling is increased~\cite{PhysRevA.102.043304}. Additionally, Ref.~\cite{PhysRevA.102.053308} showed that beyond-mean-field fluctuations can stabilize the SO-coupled ST droplet in a three-dimensional (3D) gas and constructed an approximate energy functional for the Lee--Huang--Yang (LHY) energy for finite-system calculations. Related studies have explored quantum solitons in one-dimensional (1D) SO-coupled mixtures~\cite{PhysRevA.99.063618}, droplets in a 3D gas subject to 1D SO-coupling~\cite{Xiong_2021}, the dynamics and spectrum of 1D SO-coupled localized states~\cite{PhysRevA.106.063315, PhysRevA.109.013321}, and Floquet-engineered 1D droplets~\cite{vhpp-2z36}.

This manuscript provides a consistent account of the stationary and traveling droplet states of 1D SO-coupled Bose-Bose mixtures, their collective excitations below the particle-emission threshold, and their stability within the framework of the local one-dimensional beyond-mean-field correction proposed by Petrov and Astrakharchik~\cite{petrov_16}.
The finite size of the system makes this analysis nontrivial. As the stripe period and droplet size vary, stationary-state configurations with different numbers of stripes exchange their energetic ordering. Various stripe configurations can coexist as stationary solutions in SO-coupled condensates~\cite{PhysRevA.107.053302}. Competing metastable states with distinctive density patterns have also been reported in dipolar quantum gases~\cite{PhysRevA.96.053630, PhysRevResearch.6.023023} and in droplet clusters in binary Bose-Einstein condensates (BECs) in 2D geometries~\cite{PhysRevLett.122.193902}. The energies of the competing stationary states determine the ground state, while their individual excitation spectra dictate how each stationary state responds to a perturbation. This distinction raises an important question: does a sudden jump in the excitation spectrum of the lowest-energy solution indicate an instability or a phase transition, or does it reflect a change in the underlying stationary state of the system whose modes are being measured? 

Motion poses a second question as SO-coupling breaks Galilean invariance~\cite{PhysRevA.87.063610, PhysRevA.87.013614}. As a result, the commonly employed technique of phase
imprinting on a stationary SO-coupled droplet does not, in general, yield a corresponding traveling steady-state solution \cite{PhysRevA.87.063610, PhysRevA.87.013614}. Traveling supersolid stripe patterns and
supercurrent-carrying supersolids have been studied in uniform condensates~\cite{wf3t-h862, PhysRevResearch.6.023048}, and direction-dependent self-binding was recently predicted for a detuned three-dimensional mixture with SO coupling~\cite {8nvh-541r}. 
Additionally, the merging and splitting of quantum droplets during collisions have been studied theoretically for $1\text {D}$ \cite{PhysRevA.98.013631} and both theoretically and experimentally for $3\text{D}$ Bose-Bose mixtures ~\cite{PhysRevLett.122.090401} without SO coupling.

In this work, we begin by employing a variational framework to explore the energetic competition among the ST, PW, and ZM phases in
an infinitely extended SO-coupled mixture. Subsequently, we conduct numerical periodic-cell calculations for an infinite droplet to assess the accuracy of the variational approach. A finite-size droplet analysis then links this bulk reference to localized stationary states. In the investigated regime, we find a direct ST-ZM ground-state transition and calculate the spectra of both the ground and metastable states. The smooth evolution of the metastable-state spectra elucidates the discontinuities observed in the ground-state spectrum.
Then we analyze how motion affects the ground-state phases and their stability. We find that, in a comoving frame, droplets acquire a finite magnetization and that the two ground-state phases are traveling ST and PW droplets. We verify the stability of traveling droplets using real-time evolution in the space-fixed frame and Bogoliubov analysis in the comoving frame. Lastly, we demonstrate head-on binary collisions between the traveling droplets.

The rest of the paper is organized as follows. In Sec.~\ref{Sec.II}, we introduce the energy functional for a homogeneous, one-dimensional, weakly interacting SO-coupled pseudo-spinor Bose-Bose mixture. 
We then write the coupled Gross–Pitaevskii (GP) and Bogoliubov–de Gennes (BdG) equations, incorporating the LHY correction via Petrov's prescription \cite{petrov_16}. In Sec.~\ref{Sec.III}, we use the variational method to analyze the stationary-state phases of an infinitely extended SO-coupled 1D mixture and demonstrate 
its validity with numerical minimization of the extended GP energy functional. We then study the stationary states and collective excitations of a finite-size 
SO-coupled droplet in Sec.~\ref{Sec.IV}. We use the Galilean-boosted framework in Sec.~\ref{Sec.V} to study traveling droplet phases,
their stability, and collision dynamics.
Finally, we summarize and conclude our work in Sec.~\ref{VI}.

\section{Methodology}\label{Sec.II}
We consider a homogeneous 1D weakly interacting pseudo-spinor Bose--Bose mixture with an attractive first-order quantum correction, obtained under the approximation $g_{\uparrow\downarrow} \approx -\sqrt{g_{\uparrow\uparrow}g_{\downarrow\downarrow}}$ ($g_{\uparrow\uparrow}=g_{\downarrow\downarrow}=g$) following the framework of Refs.~\cite{petrov_16, petrov_notes}. 
The energy functional of such a homogeneous mixture is
\begin{equation}
\label{enfn}
\begin{split}
    E &= \int dx \Bigg[\begin{pmatrix} \Phi_{\uparrow}^* & \Phi_{\downarrow}^* \end{pmatrix} H \begin{pmatrix} \Phi_{\uparrow}  \\ \Phi_{\downarrow} \end{pmatrix} + \frac{g}{2}\left(n_{\uparrow}^2+n_{\downarrow}^2\right) \\
    &\quad + g_{\uparrow\downarrow}n_{\uparrow}n_{\downarrow} -\frac{2\sqrt{m}g^{3/2}}{3\hbar\pi}
n^{{3}/{2}} \Bigg],
\end{split}
\end{equation}
where $\Phi_i(x,t)$ (with $i=\uparrow$ or $\downarrow$) is the component wavefunction, 
$ n_i(x,t) = |\Phi_i(x,t)|^2$ is the particle density of the $i$th spin state, and $n(x,t) = \sum_in_i(x,t)$ is the total density. 
Here, the coupling strengths are related to the 1D scattering lengths via  $g_{ij} = -2\hbar^2/(ma_{ij})$, where $n|a_{ij}|\gg 1$ ensures the system is in the weakly interacting regime~~\cite{petrov_16, PhysRevLett.122.105302}.
The parameters 
$g_{\uparrow \uparrow} = g_{\downarrow \downarrow} = g$ and $g_{\uparrow\downarrow}$ represent the intra- and inter-component coupling strengths, respectively.
The single-particle Hamiltonian is given by
\begin{align}
\label{ham}
    H = \frac{1}{2m}\left(-i \hbar \frac{\partial}{\partial x} - \hbar k_R \sigma_z\right)^2 + \frac{\hbar \Omega}{2} \sigma_x,
\end{align}
where $m$ is the atomic mass, $k_R$ characterizes the strength of the spin-orbit coupling, $\Omega$ is the Raman coupling strength, and $\sigma_{x,z}$ are the Pauli matrices.
The last term in Eq.~(\ref{enfn}) corresponds to the LHY correction. In the self-trapped droplet or soliton regime, the bulk density can be sufficiently large that the characteristic interaction energy per 
particle far exceeds the recoil and Rabi energy scales, i.e., $gn\gg E_R = \hbar^2k_R^2/2m\sim \hbar\Omega$. In this regime, the effects of spin-orbit and Rabi coupling on the leading-order quantum-
fluctuation correction can be neglected, and we approximate the LHY correction for the SO-coupled mixture in Eq.~(\ref{enfn}) by that of the corresponding mixture without SO coupling. 
The coupled GP equations describing the dynamics of the system are    
\begin{subequations}\label{gpe}
\begin{align}
i\hbar\frac{\partial\Phi_{\uparrow}}{\partial t}  =& \Bigg[ -\frac{\hbar^2}{2m}\frac{\partial^{2}}{\partial x^2} +i \frac{\hbar^2 k_{R}}{m}\frac{\partial}{\partial x}+ \frac{\hbar^2 k_{R}^2}{2m}+ g n_{\uparrow}+g_{\uparrow\downarrow}n_{\downarrow} \nonumber\\ 
& + \Delta\mu_\uparrow \Bigg]\Phi_{\uparrow} +\frac{\hbar\Omega}{2} \Phi_{\downarrow},\\
i\hbar\frac{\partial\Phi_{\downarrow}}{\partial t} =& \Bigg[ -\frac{\hbar^2}{2m}\frac{\partial^{2}}{\partial x^2}-i \frac{\hbar^2 k_{R}}{m}\frac{\partial}{\partial x}+ \frac{\hbar^2 k_{R}^2}{2m} + g n_{\downarrow}+g_{\uparrow\downarrow}n_{\uparrow} \nonumber \\
& + \Delta\mu_\downarrow\Bigg]\Phi_{\downarrow} +\frac{\hbar\Omega}{2} \Phi_{\uparrow},
\end{align}
\end{subequations}
where $\Delta\mu_{i} = -g^{3/2}\sqrt{m n}/\pi\hbar$ is the LHY correction term. We cast the equations into a dimensionless form by expressing the length, time, energy, and wavefunctions in units of $\hbar^2/(mg)$, $\hbar^3/mg^2$, $mg^2/ \hbar^2$, and $\sqrt{mg}/\hbar$, respectively. Accordingly, the inter-component coupling constant $g_{\uparrow\downarrow}$, the recoil momentum $k_R$ and Rabi frequency $\Omega$ are expressed in units of $g, mg/\hbar^2$, and $mg^2/{\hbar^3}$, respectively. The resulting dimensionless GP equations are given as
\begin{subequations}\label{scaled_gpe}
\begin{align}
i\frac{\partial \Phi_{\uparrow}}{\partial t}
=&
\Bigg[
-\frac{1}{2}\frac{\partial^{2}}{\partial x^{2}}
+i k_{R}\frac{\partial}{\partial x}
+\frac{k_{R}^{2}}{2}
+|\Phi_{\uparrow}|^{2}
+g_{\uparrow\downarrow}|\Phi_{\downarrow}|^{2}
\nonumber \\&-\frac{1}{\pi}
\sqrt{
|\Phi_{\uparrow}|^{2}
+|\Phi_{\downarrow}|^{2}
}
\Bigg]\Phi_{\uparrow}
+\frac{{\Omega}}{2}\Phi_{\downarrow},
\\
i\frac{\partial \Phi_{\downarrow}}{\partial t}
=&
\Bigg[
-\frac{1}{2}\frac{\partial^{2}}{\partial x^{2}}
-i k_{R}\frac{\partial}{\partial x}
+\frac{k_{R}^{2}}{2}
+|\Phi_{\downarrow}|^{2}
+g_{\uparrow\downarrow}|\Phi_{\uparrow}|^{2}
\nonumber \\&-\frac{1}{\pi}
\sqrt{
|\Phi_{\uparrow}|^{2}
+|\Phi_{\downarrow}|^{2}
}
\Bigg]\Phi_{\downarrow}
+\frac{{\Omega}}{2}\Phi_{\uparrow}.
\end{align}
\end{subequations}
To calculate the excitation spectrum, we linearize Eq.~(\ref{scaled_gpe}) by considering fluctuations 
around the ground state $\phi_{i}(x)$ using 
\begin{equation}
    \Phi_i(x,t) = e^{-i \mu t}\Big[\phi_i(x) + \delta{\phi}_{i}(x,t)\Big],
\end{equation}
where $\mu$ is the chemical potential. Then, using the Bogoliubov transformation, 
\begin{align}
    \delta{\phi}_{i}(x,t)=\sum_{l}\left[u_i^l (x) e^{-i \omega_l t}+v_i^{l*}(x)e^{i\omega_l t} \right] \label{fluc},
\end{align}
we obtain the coupled BdG equations, which in matrix form read as
\\
\begin{widetext}
\begin{equation}\label{bdg}
\begin{split}
\begin{pmatrix}
 \mathcal{M} & (1+\alpha)\phi_\uparrow^2  & (g_{\uparrow\downarrow}+\alpha)\phi_\uparrow\phi_\downarrow^* + \Omega/2  & (g_{\uparrow\downarrow}+\alpha)\phi_\uparrow\phi_\downarrow  \\[2pt] 
 -(1+\alpha)\phi_\uparrow^{*2} & -\mathcal{M}^* & -(g_{\uparrow\downarrow}+\alpha)\phi_\uparrow^*\phi_\downarrow^* & -(g_{\uparrow\downarrow}+\alpha)\phi_\uparrow^*\phi_\downarrow - \Omega/2 \\[2pt] 
 (g_{\uparrow\downarrow}+\alpha)\phi_\uparrow^*\phi_\downarrow + \Omega/2 & (g_{\uparrow\downarrow}+\alpha)\phi_\uparrow\phi_\downarrow & \mathcal{N} & (1+\alpha)\phi_\downarrow^2 \\[2pt] 
-(g_{\uparrow\downarrow}+\alpha)\phi_\uparrow^*\phi_\downarrow^* & -(g_{\uparrow\downarrow}+\alpha)\phi_\downarrow^*\phi_\uparrow - \Omega/2  & -(1+\alpha)\phi_\downarrow^{*2} & -\mathcal{N}^*
\end{pmatrix}
\begin{pmatrix}
 u_\uparrow^l \\[2pt]  v_\uparrow^l \\[2pt] u_\downarrow^l \\[2pt] v_\downarrow^l
\end{pmatrix}
&= \omega_l
\begin{pmatrix}
 u_\uparrow^l \\[2pt] v_\uparrow^l \\[2pt] u_\downarrow^l \\[2pt] v_\downarrow^l
\end{pmatrix},
\end{split}
\end{equation}
\end{widetext}
with
\begin{subequations}
    \begin{align}
        \mathcal{M} &= -\frac{1}{2}\frac{\partial^2}{\partial x^2}+i k_{R}\frac{\partial}{\partial x}+ \frac{k_{R}^2}{2}+  2|\phi_\uparrow|^2+g_{\uparrow\downarrow}|\phi_\downarrow|^2 \nonumber \\& \qquad  -\mu +\alpha \left( 3|\phi_\uparrow|^2 +2|\phi_\downarrow|^2\right), \\ 
        \mathcal{N} &= -\frac{1}{2}\frac{\partial^2}{\partial x^2}  -i k_{R}\frac{\partial}{\partial x}+ \frac{k_{R}^2}{2}+ 2|\phi_\downarrow|^2+g_{\uparrow\downarrow}|\phi_\uparrow|^2 \nonumber \\& \qquad -\mu +\alpha \left(3|\phi_\downarrow|^2 +2|\phi_\uparrow|^2\right),
    \end{align}
\label{matrix_element}
\end{subequations}
where $\alpha = -1/2\pi\sqrt{n}$.

\section{Variational Analysis}\label{Sec.III} 
To understand the ground-state properties, such as the energy per particle, bulk density, and momentum composition of large droplets, we first examine the ground-state phases of an 
infinitely extended SO-coupled 1D liquid.
For this purpose, we consider the following variational \textit{ansatz}~\cite{PhysRevLett.108.225301}
\begin{align}
\label{par}
\begin{bmatrix}
\Phi_{\uparrow}(x) \\  \Phi_{\downarrow}(x) \end{bmatrix} = \sqrt{n_0}
\Bigg[c_1\begin{pmatrix}
\cos{\theta} \\ -\sin{\theta} \end{pmatrix} e^{i k_1 x} \nonumber\\ +c_2\begin{pmatrix}
\sin{\theta} \\ -\cos{\theta} \end{pmatrix} e^{-i k_1 x}\Bigg],
\end{align}
where $n_0$, $c_1, c_2, k_1,$ and $\theta$ are variational parameters with $|c_1|^2+|c_2|^2=1$. 
Using the variational {\em ansatz}~(\ref{par}) in Eq.~(\ref{enfn}), we obtain the variational energy as a function of $n_0, k_1, \beta = |c_1|^2|c_2|^2$, and $\theta$. 
The stationarity condition with respect to $k_1$ yields
$k_1=k_R\cos(2\theta)$,
which allows us to eliminate $\theta$ in favor of $k_1$: $\theta =\arccos{(k_1/k_R)}/2$.
The total density $n(x)$ and the longitudinal and transverse spin polarization $\langle\sigma_z\rangle$ and $\langle \sigma_x\rangle$ can then be written as
\begin{align}
n(x) &= n_0\left(1+2|c_1||c_2|\sqrt{1-\frac{k_1^2}{k_R^2}}\cos(2k_1x)\right),\\
\langle\sigma_z\rangle&=  \frac{1}{N_{l}}\int_0^{l} dx \left(\Phi_{\uparrow}^*, \Phi_{\downarrow}^*\right)\sigma_z \left(\Phi_{\uparrow}, \Phi_{\downarrow}\right)^T \nonumber\\ 
&= k_1\frac{|c_1|^2-
|c_2|^2}{k_R},\label{sigmaz}\\
\langle\sigma_x\rangle&= \frac{1}{N_{l}}\int_0^{l} dx \left(\Phi_{\uparrow}^*, \Phi_{\downarrow}^*\right)\sigma_x \left(\Phi_{\uparrow}, \Phi_{\downarrow}\right)^T\nonumber\\
&=-\sqrt{1-\frac{k_1^2}{k_R^2}},
\label{sigmax}
\end{align}
where $l = 2\pi/k_1$ is the period of the variational \textit{ansatz} (for $k_1 \ne 0$), $ N_l = \int_0^l n(x) dx$,  $n_0 = N_l/l$ (a variational parameter) is the average density 
and the phase difference between $c_1$ and $c_2$ has been assumed to be zero.

Using the variational {\em ansatz} [Eq.~(\ref{par})] in Eq.~(\ref{enfn}), the (dimensionless) variational energy per particle is
\begin{align}\label{var_E}
\mathcal{E} = &\frac{1}{N_l}\int_0^{l} dx \Bigg[\left(\Phi_{\uparrow}^*, \Phi_{\downarrow}^*\right)H \left(\Phi_{\uparrow}, \Phi_{\downarrow}\right)^T + \frac{1}{2}
\left(n_{\uparrow}^2+n_{\downarrow}^2\right)\nonumber \\
    &\quad + g_{\uparrow\downarrow}n_{\uparrow}n_{\downarrow} -\frac{2}{3\pi}n^{3/2}\Bigg].
\end{align}
To obtain the contribution of the LHY correction to the variational energy, Eq.~(\ref{var_E}), which is $\propto \int n(x)^{3/2} dx$,
we approximate  $n(x)^{3/2}$ as follows:
\begin{align}
    n(x)^{3/2}=&n_0^{3/2}\left(1+2 \sqrt{\beta} \sqrt{1-\frac{k_1^2}{k_R^2}} \cos (2 k_1 x)\right)^{3/2} \nonumber \\ \simeq&  n_0^{3/2} \Bigg[1+3\sqrt{\beta} \sqrt{1-\frac{k_1^2}{k_R^2}} \cos (2 k_1 x) \nonumber\\+&\frac{3}{2}\beta \left(1-\frac{k_1^2}{k_R^2}\right) \cos^2(2 k_1 x)\Bigg],
\end{align}
assuming 
\begin{equation}
2 \frac{\sqrt{\beta}}{k_R} \sqrt{k_R^2-k_1^2}\ll1.\label{lhy_exp} 
\end{equation}
The total variational energy per particle thus obtained is
\begin{align}
\label{enpp}
    \mathcal{E}&\simeq \frac{k_R^2}{2}-\frac{f(\beta,n_0) k_1^2}{2k_R^2}+(2 \beta +1) G_1-\frac{\Omega}{2k_R}\sqrt{k_R^2-k_1^2} \nonumber\\&-\frac{\sqrt{n_0}}{2\pi}\left(\frac{4}{3}+\beta\right), 
\end{align}
where $G_1=n_0(1+g_{\uparrow\downarrow})/4$, $G_2=n_0(1-g_{\uparrow\downarrow})/4$, and 
\begin{equation}\label{f_beta}
f(\beta,n_0) = k_R^2-2G_2+4\beta(G_1+2G_2)-\beta\frac{\sqrt{n_0}}{\pi},
\end{equation}
which is linear in $\beta$ with $\beta\in[0,1/4]$. We assume $\Omega\ge0$ and $k_1\ge 0$ without loss of generality. In the remainder
of this section, we consider $-1<g_{\uparrow\downarrow}<1$ and $k_R>0$.

Minimizing Eq.~(\ref{enpp}) with respect to $k_1$, 
we get
\begin{equation}\label{k1_var}
k_1 = 
\begin{cases}
    0 & \text{if } \Omega\ge 2f(\beta,n_0),  \\
     k_R\sqrt{1-\frac{\Omega ^2}{4 f(\beta,n_0)^2}} & \text{if } 0 \le \Omega<2f(\beta,n_0).
\end{cases}
\end{equation}

\subsection{Zero-momentum droplet}
For $\Omega\ge2f(\beta,n_0)$, $k_1= 0$, which on substituting in Eq.~(\ref{enpp}) yields 
\begin{align}
    \mathcal{E}&\simeq\frac{k_R^2}{2}+(2 \beta +1) G_1-\frac{\Omega}{2} -\frac{\sqrt{n_0}}{2\pi}\left(\frac{4}{3}+\beta\right).  
\end{align}
Minimizing it with respect to $\beta$ and $n_0$, simultaneously, for $g_{\uparrow\downarrow}>-1$ gives 
$\beta = 0$ and the saturation density for the zero momentum (ZM) droplet
\begin{equation}
n_{\rm ZM}^{\rm s} \simeq \left[\frac{4}{3\pi(1+g_{\uparrow\downarrow})}\right]^2.
\end{equation}
The corresponding minimum energy per particle is given as 
\begin{equation}
\mathcal{E}_{\rm ZM}^{\rm droplet}\simeq -\frac{4}{9 \pi ^2 (g_{\uparrow\downarrow}+1)}-\frac{\Omega }{2}+\frac{k_R^2}{2}, \label{E_ZM}
\end{equation}
where
\begin{subequations} \label{ZM_cond}
\begin{align}g_{\uparrow\downarrow}>&-1\quad{\rm and}\\
\Omega\ge&2f(0,n_{\rm ZM}^{\rm s})\nonumber\\
         =& 2\left[k_R^2-\frac{8  (1-g_{\uparrow\downarrow})}{9 \pi ^2(g_{\uparrow\downarrow}+1)^2}\right]. 
\end{align}
\end{subequations}
As $\beta = 0$, inequality (\ref{lhy_exp}) is trivially satisfied. Alternatively, for the zero-momentum phase, the variational ansatz has a simpler form,
 \begin{align}
    \begin{bmatrix}
 \Phi_{\uparrow}(x) \\  \Phi_{\downarrow}(x) \end{bmatrix} = \sqrt{n_0}
\Bigg[\begin{pmatrix}
 \cos{\theta} \\ -\sin{\theta} \end{pmatrix} \Bigg] \label{ZM_anasatz}.
\end{align}
Substituting the above {\em ansatz} in Eq.~(\ref{enfn})  and minimizing the 
energy with respect to $\theta$ gives $\theta = \pi/4$. The resultant variational energy per particle is
\begin{equation}
\mathcal{E}_{\rm ZM}^{\rm droplet} = -\frac{4}{9 \pi ^2 (g_{\uparrow\downarrow}+1)}-\frac{\Omega }{2}+\frac{k_R^2}{2},
\end{equation}
which on minimization with respect to $n_0$, gives
\begin{equation}
n_{\rm ZM}^{\rm s} = \left[\frac{4}{3\pi(1+g_{\uparrow\downarrow})}\right]^2.
\end{equation}

\subsection{Finite momentum droplets}
For $0\le \Omega<2f(\beta,n_0)$, $ k_1= k_R\sqrt{1-\Omega ^2/4 f(\beta,n_0)^2}\ne0$, which when inserted in
Eq.~(\ref{enpp}) gives 
\begin{align}
    \mathcal{E}\simeq &\frac{k_R^2-f(\beta,n_0)}{2}+(2 \beta +1) G_1-\frac{\Omega^2}{8 f(\beta,n_0)} \nonumber\\
               & -\frac{\sqrt{n_0}}{2\pi}\left(\frac{4}{3}+\beta\right) \nonumber\\
                =&  G_1 + G_2(1-4\beta) - \frac{2\sqrt{n_0}}{3\pi}-\frac{\Omega^2}{8 f(\beta,n_0)}. \label{energy_finite_m} 
\end{align}
The second derivative of energy per particle
\begin{align}
    \frac{\partial^2\mathcal{E}}{\partial \beta^2}&= \begin{cases} 
  -\frac{\Omega^2}{4 f^3}\left(\frac{\partial f}{\partial \beta}\right)^2<0 & \text{if }  0< \Omega<2f(\beta,n_0)\\
  0 & \text{if } \Omega=0.
\end{cases} 
\end{align}
The energy in Eq.~(\ref{energy_finite_m}) is a concave function in $\beta$ for $0< \Omega<2f(\beta,n_0)$, which implies that the minimum over
a closed interval $\beta\in[0,1/4]$ occurs at one of the end points, $\beta = 0$ or $1/4$. For $\Omega = 0$, energy 
becomes a linear function of $\beta$ with a negative slope and a minimum at $\beta = 1/4$. 

Again for $\beta = 0$, the inequality (\ref{lhy_exp}) holds trivially true, whereas substituting $\beta = 1/4$ and $k_1 = k_R\sqrt{1-\frac{\Omega ^2}{4 f(\beta,n_0)^2}}$
in inequality (\ref{lhy_exp}) yields $0\le\Omega\ll2f(1/4,n_0)$.

\subsubsection*{Plane wave droplet with $\beta = 0, k_1\ne0$} 
Here $f(\beta= 0,n_0) =  k_R^2-2G_2 >0$ for $0<n_0<2k_R^2/(1-g_{\uparrow\downarrow})$, and the energy per particle of the PW phase 
is
\begin{align}
    \mathcal{E_{\rm PW}}(n_0)&\simeq G_1 + G_2 - \frac{2\sqrt{n_0}}{3\pi}-\frac{\Omega^2}{8 f(0,n_0)}.\label{epw}
\end{align}
For the PW phase with $n_0 \ll 2k_R^2/(1-g_{\uparrow\downarrow})$, which satisfies the weaker condition for $f(0,n_0)$ to be positive, the minima of $\mathcal{E_{\rm PW}}(n_0)$ can be estimated by approximating it as 
\begin{equation}
    \mathcal{E_{\rm PW}}(n_0)\simeq G_1 + G_2 - \frac{2\sqrt{n_0}}{3\pi}-\frac{\Omega^2}{8 k_R^2}\left(1+\frac{2G_2}{k_R^2}\right),
\end{equation}
near the vicinity of minima. 
The minimization leads to
\begin{align}
n_{\rm PW}^{\rm s}\simeq& \frac{256}{9 \pi ^2 \left[(g_{\uparrow\downarrow}-1) \Omega ^2/k_R^4+8\right]^2},\\
\mathcal{E}_{\rm PW}^{\rm droplet} \simeq&-\frac{16}{9 \pi ^2 \left[(g_{\uparrow\downarrow}-1) \Omega ^2/ k_R^4+8\right]}-\frac{\Omega ^2}{8 k_R^2},\label{E_PW}
\end{align}
provided 
\begin{subequations} \label{PW_cond}
\begin{equation}
0\le\Omega < \frac{2\sqrt{2}k_R^2}{\sqrt{1-g_{\uparrow\downarrow}}};
\end{equation}
additionally
\begin{align}
n_{\rm PW}^{\rm s}\ll&2k_R^2/(1-g_{\uparrow\downarrow}), \quad {\rm and}\\ 
0\le\Omega <& 2f(0,n_{\rm PW}^{\rm s}).
\end{align}
\end{subequations}
\subsubsection*{Stripe droplet with $\beta = 1/4, k_1\ne0$} 
The energy per particle of the ST phases 
is
\begin{align}
    \mathcal{E_{\rm ST}}(n_0)&\simeq G_1 - \frac{2\sqrt{n_0}}{3\pi}-\frac{\Omega^2}{8 f(1/4,n_0)},
\end{align}
where $f(1/4,n_0)= k_R^2+G_1-\sqrt{n_0}/4\pi >0$ for all $n_0>0$, provided $1+g_{\uparrow\downarrow}>1/(16\pi^2k_R^2)$.
For $|G_1-\sqrt{n_0}/(4\pi)|\ll k_R^2$ 
the energy per particle can be approximated as
\begin{equation}
    \mathcal{E_{\rm ST}}(n_0)\simeq G_1 - \frac{2\sqrt{n_0}}{3\pi}-\frac{\Omega^2 }{8k_R^2}\left(1-\frac{G_1}{k_R^2}+\frac{\sqrt{n_0}}{4\pi k_R^2}\right).
\end{equation}
Minimizing $\mathcal{E}_{\rm ST}$ with respect to $n_0$ gives the saturation density,
\begin{equation}
n_{\rm ST}^{\rm s} \simeq \frac{\left(3 \Omega ^2+64k_R^4\right)^2}{36 \pi ^2 \left(\Omega ^2+8k_R^4\right)^2 (1+g_{\uparrow\downarrow})^2},\label{n_st}
\end{equation}
and the energy per particle
\begin{align}
 \mathcal{E}_{\rm ST}^{\rm droplet}\simeq&   -\frac{1}{1152 \pi ^2 (g_{\uparrow\downarrow}+1)  \left(8 +\Omega ^2/k_R^4\right)}\nonumber\\&\times\Bigg[9 \frac{\Omega ^4}{k_R^8} \left(16 \pi ^2 (g_{\uparrow\downarrow}+1)k_R^2 +1\right)\nonumber\\&+384  \frac{\Omega ^2}{k_R^4} \left(3 \pi ^2 (g_{\uparrow\downarrow}+1)k_R^2 +1\right)+4096 \Bigg] \label{E_ST}
\end{align}
of the ST droplet, where 
\begin{subequations}\label{ST_cond}
\begin{equation}
1+g_{\uparrow\downarrow}>\frac{1}{16\pi^2k_R^2},
\end{equation} 
in addition to 
\begin{align}
\left|\frac{n_{\rm ST}^{\rm s}}{4}(1+g_{\uparrow\downarrow})-\frac{\sqrt{n_{\rm ST}^{\rm s}}}{4\pi}\right|\ll& k_R^2, \quad {\rm and}\\
0\le\Omega\ll2f(1/4,n_{\rm ST}^{\rm s})&.
\end{align}
\end{subequations}

\begin{figure}
\centering
\includegraphics[width=0.9\columnwidth]{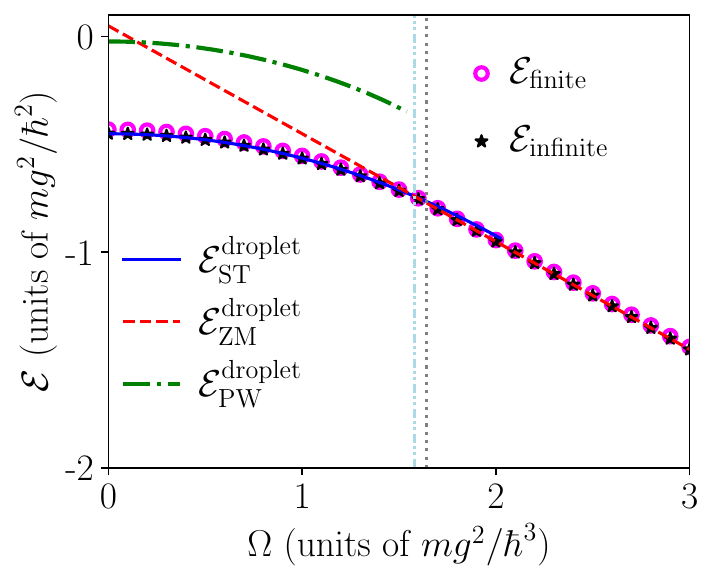}
\caption{
Energy per particle $\mathcal{E}$ as a function of the Rabi coupling strength $\Omega$ for a droplet with $g_{\uparrow \downarrow} = -0.9$. 
Solid black stars represent the numerical ground-state energy per particle for the infinite-size system, while open magenta circles show the corresponding results 
for a finite-size droplet with $N = 300$. 
The lines show variational estimates of $\mathcal{E}$ given by Eq.~(\ref{E_ZM}) (dashed red), Eq.~(\ref{E_PW}) (dash-dotted green), and Eq.~(\ref{E_ST}) (solid blue). As discussed in the text, 
$\mathcal{E}$ obtained by the variational method is plotted over the ranges $0 \leq \Omega < 2$ for the ST phase and $0 \leq \Omega < 1.54$ for the PW phase. The vertical lines at $\Omega_{\rm cr}^{\rm var}=1.58$ (light blue) and $\Omega_{\rm cr}^{\rm num}=1.64$ (gray) indicate the boundary between the ST and ZM phases for an infinite-size droplet, using variational and numerical calculations, respectively.}
\label{energy_plot} 
\end{figure}
\begin{figure}
\centering
\includegraphics[width=\columnwidth]{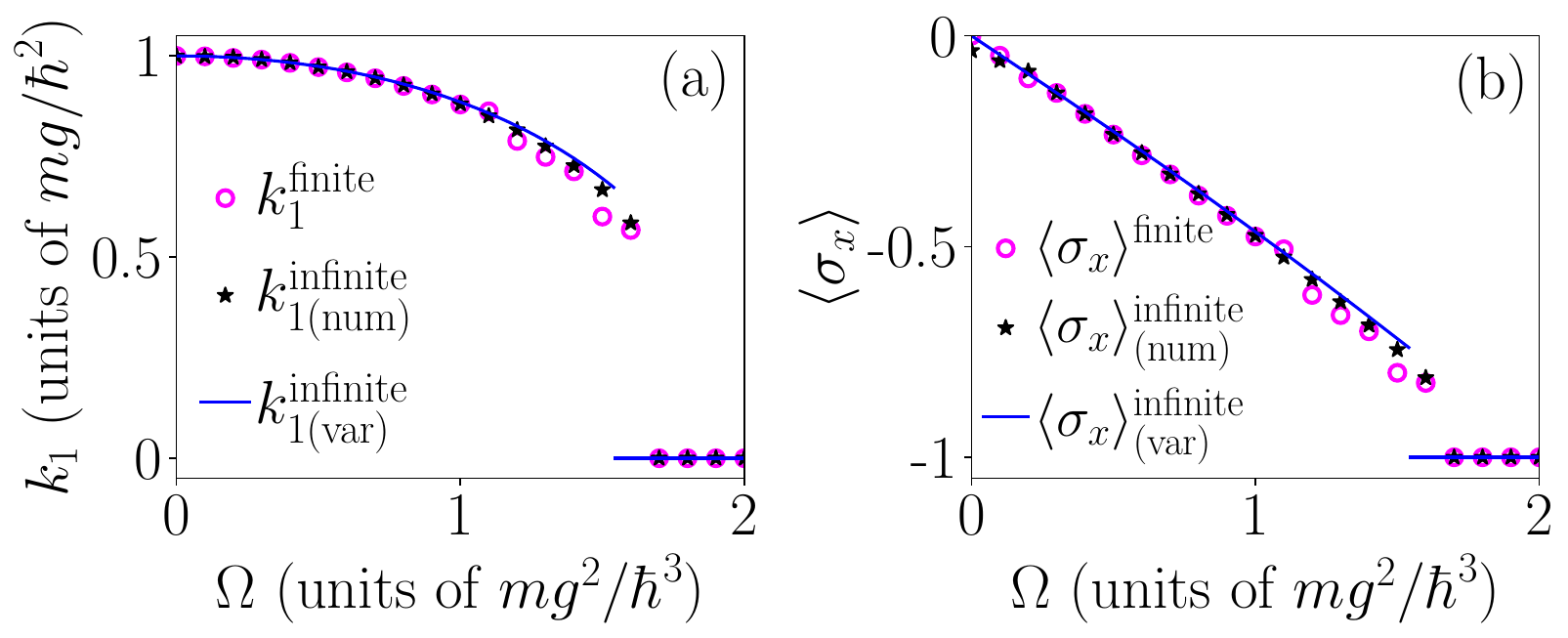}
\caption{(a) Condensate momentum $k_1$ and (b) transverse spin polarization $\langle\sigma_x\rangle$ as a function of the Raman coupling $\Omega$. The solid blue line represents the results from the variational approach for the infinite droplet, whereas the symbols represent the numerically obtained values for the infinite droplet (black stars) and the finite-size droplet (open magenta circles). Here, we consider $N=300$ and $g_{\uparrow \downarrow}=-0.9$. The differences close to the ST-ZM boundary are due to higher-order harmonics and finite-size effects.}
\label{k1_sigmax} 
\end{figure}

To validate the variational analysis, we numerically evaluate the ground state properties of an infinitely extended SO-coupled 1D liquid.
We briefly describe the numerical procedure to simulate an SO-coupled 1D liquid in the thermodynamic limit. 
We consider the spatial period of the order parameter, $l$, and the number of atoms per period, $n_0=N_l/l$, as the variable input parameters.
We solve the coupled GP equations (\ref{scaled_gpe}) for a range of $(n_0,l)$ values using imaginary time propagation. 
The variation in $n_0$ and $l$ is considered over suitably chosen ranges to ensure that a minimum of energy per particle lies 
within the truncated $(n_0,l)$ plane.
We consider $\Delta n_0 = 0.05$ and $\Delta l = 0.01$ as the step sizes to vary the average density and 
spatial extent. For each pair of $(n_0,l)$ values, we consider a spatial step size $\Delta x = l/1024$ and an imaginary time step size 
$\Delta \tau = 0.1\times\Delta x^2$ to solve the GP equations in imaginary time using the Fourier pseudospectral method \cite{kaur2021fortress, *banger2022fortress, *banger2021semi}. 
This gives us energy as a function of $(n_0, l)$  whose minimum fixes the saturation density and period of the wavefunction.
The ground state energy per particle $\mathcal {E}$ obtained numerically using the procedure
for an infinitely extended SO-coupled Bose gas with $k_R =1$ and $g_{\uparrow\downarrow}=-0.9$ 
as a function of $\Omega$ is plotted (using solid black stars) in Fig. \ref{energy_plot}, which shows that the energy per particle  
decreases with an increase in the Raman coupling strength. The numerically obtained $k_1$ and $\langle \sigma_x \rangle $ 
as a function of $\Omega$ are shown in Figs.~\ref{k1_sigmax}(a) and \ref{k1_sigmax}(b), respectively.
For $0<\Omega<\Omega_{\rm cr}^{\rm num}$, the SO-coupled liquid is in the ST phase with a transition from the ST to the 
ZM phase at $\Omega_{\rm cr}^{\rm num}\approx 1.64 $. The energy decreases linearly with $\Omega$ above $\Omega>\Omega_{\rm cr}^{\rm num}$. 
Both $k_1$ and $\sigma_x$ decrease with an increase in $\Omega$ in 
the stripe phase and change discontinuously across the critical point to become zero and $-1$, respectively, in the zero-momentum phase [see Figs.~\ref{k1_sigmax}(a) and \ref{k1_sigmax}(b)].

Before we compare numerical results in Figs.~\ref{energy_plot} and \ref{k1_sigmax} with variational results, we first ascertain the domains of applicability of the variational results  
from the conditions (\ref{ZM_cond}), (\ref{PW_cond}), and (\ref{ST_cond}) for $k_R =1$ and $g_{\uparrow\downarrow}=-0.9$. 
In the zero-momentum phase, any non-negative $\Omega$ satisfies the inequality (\ref{ZM_cond}). 
The most restrictive condition for the validity of the variational approximation in the plane-wave phase, given by (\ref{PW_cond}c), translates to $0 \le \Omega < 1.54$. 
Similarly, for the stripe phase, the condition (\ref{ST_cond}c) leads to the most stringent requirement of $0 \le \Omega \ll 2$ for the validity of the variational estimate.
The variational energies obtained for the ZM, PW, and ST phases, viz., Eqs.~(\ref{E_ZM}), (\ref{E_PW}), and (\ref{E_ST}), respectively, are plotted using lines in Fig.~\ref{energy_plot}.
The ST and the ZM phase, respectively, emerge as the minimum energy phase 
below and above critical point $\Omega_{\rm cr}^{\rm var} = 1.58$ defined by the positive real solution of $\mathcal{E}_{\rm ST}^{\rm droplet}(\Omega) = \mathcal{E}_{\rm ZM}^{\rm droplet}(\Omega)$ in a good agreement with numerical results. The deviation of $\Omega_{\rm cr}^{\rm var}$ from $\Omega_{\rm cr}^{\rm num}$ is due to the absence of higher harmonics in the variational {\em ansatz} (\ref{par}).
The PW phase remains metastable across $0 \le \Omega < 1.54$. Similarly, the variationally obtained $k_1$ (defined as $k_1 = k_R\sqrt{1-\Omega ^2/\left[4 f(1/4,n_{\rm ST}^{\rm s})^2\right]}$ for $\Omega<\Omega_{\rm cr}^{\rm var}$,  and $k_1 = 0 $ otherwise) and  
$\langle\sigma_x\rangle$  agree well with the numerical results in Figs.~\ref{k1_sigmax}(a) and \ref{k1_sigmax}(b), respectively.

\section{Finite Size Droplet and collective excitations}\label{Sec.IV}
For a finite number of atoms $N$, the ground-state solution is obtained by solving Eqs.~(\ref{scaled_gpe}a) and (\ref{scaled_gpe}b) via imaginary-time propagation using the Fourier pseudo-spectral method \cite{kaur2021fortress, *banger2022fortress, *banger2021semi}. 
We first examine the ground-state properties of a finite-size droplet with a fixed number of atoms, $N = 300$, and inter-component coupling strength, $g_{\uparrow\downarrow} = -0.9$, as a function of $\Omega $. We maintain these parameters for all finite-size droplets discussed throughout this work, unless stated otherwise.
We implemented the numerical method on a spatial domain of extent $L = 60$, using a spatial step size $\Delta x = 0.02$, and a time step size $\Delta \tau = 0.1\Delta x^2$.
The ground state density profiles of droplets at five representative values of $\Omega$ are shown in
\begin{figure*}[ht]
\centering
\includegraphics[width=\textwidth]{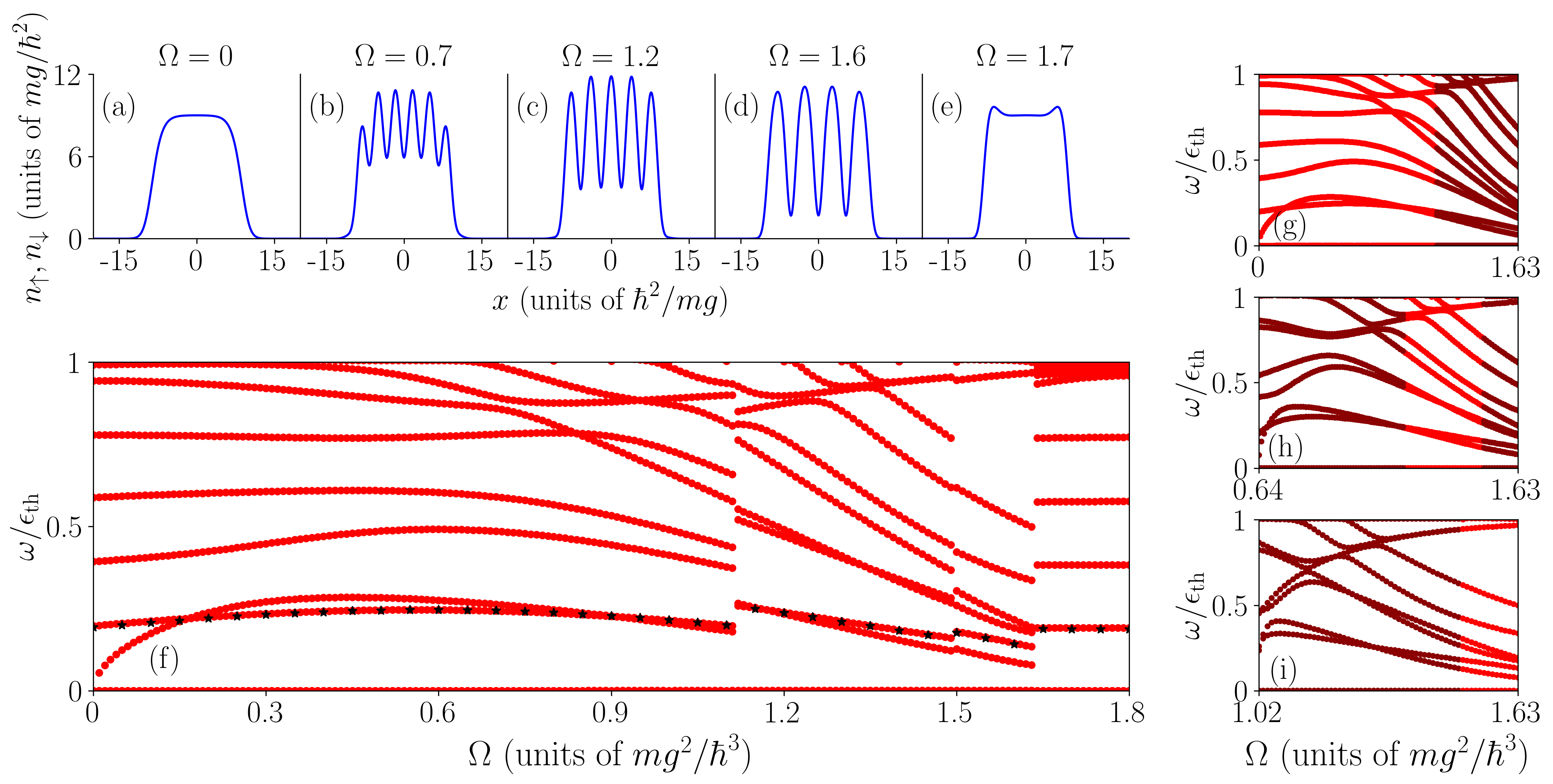}
\caption{(a)-(e) Ground-state component densities $n_{\uparrow}{=}n_{\downarrow}$ at representative Raman coupling for $N = 300$ and $g_{\uparrow \downarrow}{=}-0.9$. At $\Omega =0$, the density has zero stripe contrast; increasing $\Omega$ produces distinct stripe-number ground-state density profiles before the transition to the ZM droplet. 
(f) Excitation spectrum below the particle emission threshold as a function of $\Omega$; black stars mark the density breathing modes. Panels (g)--(i) resolve the spectra of six-, five-, and four-stripe stationary solutions, respectively, including the regions where they appear as higher-energy metastable solutions (darkred markers). The discontinuities in (f) coincide with the changes in the ground-state density configurations, whereas the frequencies for each fixed stripe number in (g)--(i) evolve smoothly.
} 
\label{mode_plot_om} 
\end{figure*}
\begin{figure*}[ht]
\centering
\includegraphics[width=0.9\textwidth]{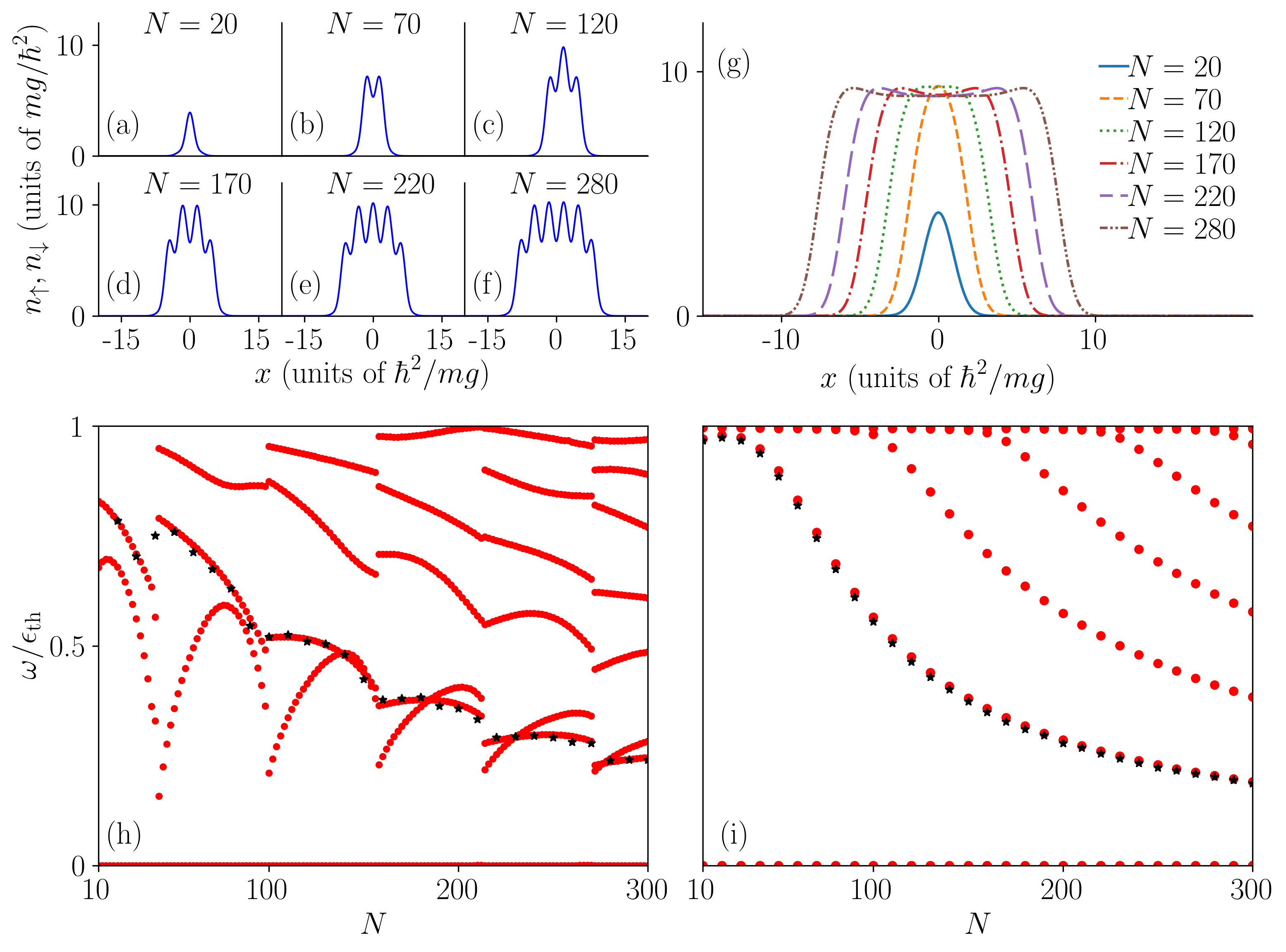}
\caption{Ground-state density profiles of quantum droplets for different numbers of atoms in (a)--(f) the stripe phase at $\Omega = 0.5$ and (g) the ZM phase at $\Omega = 2$. The excitations below the particle-emission threshold as a function of $N$ are shown for (h) $\Omega=0.5$ and (i) $\Omega=2$. Here, $g_{\uparrow \downarrow}=-0.9$. The black stars in (h) and (i) correspond to the largest peak frequencies in the Fourier transform of the oscillations in $\langle x^2\rangle$ during the real-time dynamics of the ground state generated by suitably perturbing the Hamiltonian as discussed in the text.
}
\label{mode_plot} 
\end{figure*}
Figs.~\ref{mode_plot_om}(a)--(e). As $g_{\uparrow\uparrow}=g_{\downarrow\downarrow}$, the densities of the two components overlap. For $\Omega\le1.63$, the ground state corresponds to the ST droplet, whereas for $\Omega \ge 1.64$ it transitions to the ZM droplet. The energy per particle $\mathcal {E}$ for the finite-size system as a function of $\Omega$ is
shown in Fig.~\ref{energy_plot}, along with the condensate momentum $k_1$ and the transverse spin polarization $\langle \sigma_x \rangle$ in Figs.~\ref{k1_sigmax}(a) and \ref{k1_sigmax}(b), respectively (shown by magenta circles).
These finite-size results closely approach the infinite-size droplet limit, as the droplet with $N =300$ lies well within the flat-top droplet regime. An increase in $\Omega$ leads to a decrease in $k_1$ as shown in Fig.~\ref{k1_sigmax}(a), implying an increase in the spatial period of
the stripe density pattern. As the spatial period increases, the system transitions through contiguous $\Omega$ intervals, with the number of peaks in the ground-state density profile successively decreasing by one. For instance, as $\Omega$ is increased from $0$ within the range $0<\Omega\le 1.11 $, 
the increase in the spatial period is accommodated by the ground-state solution without any change in the number of peaks in
the corresponding density profile, which features six peaks (see, e.g., Fig.~\ref{mode_plot_om}(b)). As $\Omega$ exceeds $1.11$, the number of peaks in the ground-state density profile decreases by one; consequently, for $1.12\le\Omega\le 1.49 $, the ground-state density profiles feature five peaks, as shown for a representative case in Fig. \ref{mode_plot_om}(c).
Another decrease in the number of peaks occurs above $ \Omega = 1.49$, resulting in a ground-state solution featuring four peaks until the system transitions
to the ZM droplet phase above $\Omega = 1.63$. We observe that these ground-state solutions in their respective $\Omega$ intervals continue to persist as metastable solutions 
in the contiguous $\Omega$ intervals. For example, the solution with six peaks persists as a metastable solution for $1.12\le\Omega\le 1.63$; the solution with five peaks appears as a metastable solution for $0.64\le \Omega\le 1.11$ and $1.50\le \Omega\le 1.63$, and
the one with four peaks features as a metastable solution for $1.02\le\Omega\le 1.49$.  

We also calculate the elementary excitation spectrum of the system by solving the BdG equation (\ref{bdg}) numerically.
We used a basis expansion method \cite{roy-2020, PhysRevA.111.023311} with energy
eigenstates of a particle confined to a box of length $L = 60$ serving as a suitable basis truncated to the
lowest $400$ states; low-lying eigenvalues are converged with respect to the basis size. The results were cross-checked with the finite difference 
method to solve the BdG equations \cite{GAO2020109058}. 
The ground-state excitation spectrum for $N=300$ and $g_{\uparrow \downarrow}=-0.9$ as a function of $\Omega$ is plotted
in Fig.~\ref{mode_plot_om}(f). The energies are plotted in units of particle emission threshold $\epsilon_{\rm th}$, and
only the discrete spectrum below this threshold is shown.
For the droplet pressure $P = 0$, therefore energy per particle $\mathcal{E}= E/N = \mu$. The particle emission threshold for an
SO-coupled droplet is $\epsilon_{\rm th}=\epsilon_{\rm min}-\mu $, where $\epsilon_{\rm min}$ is the minimum single-particle energy; 
$\epsilon_{\rm min} = (k_R^2-\Omega)/2$ for the ZM droplet and $\epsilon_{\rm min} = -\Omega^2/(8k_R^2)$ for the ST droplet.
We also identified the breathing mode in the ground-state excitation spectrum by examining
the time evolution of the ground state generated by suitably perturbing the Hamiltonian. Particularly, we slightly perturbed the intracomponent
couplings at $t= 0$, then examined the time evolution $\langle x^2 \rangle (t) = \int dx x^2 \left[|\phi_\uparrow(x,t)|^2+|\phi_\downarrow(x,t)|^2\right]/N $.
The dominant peak in the Fourier transform of $\langle x(t)^2 \rangle$ provides a measure of the breathing oscillation frequency, which is shown by 
the black stars in Fig.~\ref{mode_plot_om}(f).
For $\Omega\leq 1.63$ in the ST phase, the changes in the number of density peaks of the ground-state droplet are also reflected 
in the excitation spectrum as discontinuities. In Fig.~\ref{mode_plot_om}(f) at $\Omega=0$, we have three Goldstone modes (zero energy modes) 
that correspond to global phase, relative phase, and translational symmetry breakings. For $\Omega\ne0$, the relative phase between the two components
is fixed, and we get only two zero-energy modes corresponding to global phase and translational symmetry breaking. In Fig.~\ref{mode_plot_om}(f) for 
$\Omega \ge 1.64$, we have the excitation spectrum of the ZM droplet. We can see that the excitation frequencies below the particle emission threshold 
do not change with the variation in the coherent coupling for the ZM droplet.
The excitation spectra of the stripe droplets with six, five, and four peaks in their density profiles, across the $\Omega$ 
domains in which they exist as stationary states are shown in Figs. ~\ref{mode_plot_om}(g)--(i), respectively. 

We also studied the ground-state excitation spectrum by varying the number of atoms from $N=10$ to $N=300$ at $g_{\uparrow\downarrow}=-0.9$ in the ST ($\Omega = 0.5$) and ZM ($\Omega = 2$) droplet phases. The ground state density profiles in the ST and ZM droplet phases with the variation of $N$ are shown in Figs.~\ref{mode_plot}(a)--(f) and Fig.~\ref{mode_plot}(g), respectively. Initially, both the peak density and width of the droplet increase with an increase in $N$ until the peak density saturates at around  $N\approx110$ for ST and $N\approx90$ for the ZM droplet phase. A subsequent increase in the number of atoms leads to the droplet's expansion without an increase in peak density as seen in Figs.~\ref{mode_plot}(c)--(f) for the ST and Fig.~\ref{mode_plot}(g) for the ZM phase. 
The ground-state excitation spectra as a function of $N$ for  $g_{\uparrow\downarrow} = -0.9$ in the ST phase at $\Omega = 0.5$ and the ZM phase at $\Omega = 2$  
are plotted in Figs.~\ref{mode_plot}(h) and (i),
respectively, where the black stars indicate the breathing mode frequencies extracted from real-time dynamics. As $N$ is increased in the ST phase, the number of peaks in the droplet's density profile increases from one to six in discrete steps of one 
across critical $N$ values. This structural change in the stripe droplet results in discontinuities in the excitation frequencies as a function of $N$
as seen in Fig.~\ref{mode_plot}(h). We can see that as the number of atoms increases, a larger number of excitation frequencies lie below the particle emission threshold for both the ST and ZM droplet phases, consistent with observations made for a pseudospinor mixture with SO coupling \cite{PhysRevA.101.051601,w68w-68w5}. The lower-lying spectrum
of the ZM droplet is qualitatively similar to that of a mixture without SO coupling~\cite{PhysRevA.98.013631, PhysRevA.101.051601, w68w-68w5}. In the ZM phase, the out-of-phase (spin branch) is gapped by the Rabi coupling~\cite{recati_21}. As a result, these modes lie above the particle-emission threshold, leaving only in-phase density modes which are governed by the same mean-field and LHY interactions as in the absence of SO-coupling.

So far, we have focused on the ground-state properties of the static SO-coupled droplets, whose center of mass remains fixed in space with
respect to the laboratory frame. It is therefore natural to ask how these properties are modified when the droplets acquire a finite velocity. 
In the following section, we address this by discussing the ground-state properties of these droplets in a comoving frame fixed at their center of mass.

\section{Moving Droplets}\label{Sec.V}
To obtain the moving droplets with velocity $v$ in the laboratory frame, we first examine the form of the GP equations in a comoving frame of the droplet. 
We require \(vk_R\ll gn\), so that the velocity-dependent contribution is small compared to the characteristic interaction energy per particle and does not modify the LHY correction.
Using the Galilean transformation $x=x'+vt', t=t'$, which yields
\begin{align*}
    \frac{\partial}{\partial x} = \frac{\partial}{\partial x'}, \frac{\partial}{\partial t} = \frac{\partial}{\partial t'} - v\frac{\partial}{\partial x'}
\end{align*}
and
\begin{align}\label{phi_transf}
    \Phi_i(x,t) = e^{i (v x' + v^2t'/2)}\Phi_i'(x',t')
\end{align}
in Eq.~(\ref{scaled_gpe}), the GP equations in droplet's frame are
\begin{subequations}\label{tranform_gpe}
\begin{align}
i\frac{\partial \Phi'_{\uparrow}}{\partial t'}
=&
\Bigg[
-\frac{1}{2}\frac{\partial^{2}}{\partial x'^{2}}
+i k_{R}\frac{\partial}{\partial x'}
+\frac{k_{R}^{2}}{2}
+|\Phi'_{\uparrow}|^{2}
+g_{\uparrow\downarrow}{|\Phi'_{\downarrow}|^{2}}
\nonumber \\&-\frac{1}{\pi}
\sqrt{
|\Phi'_{\uparrow}|^{2}
+|\Phi'_{\downarrow}|^{2}
}
-v k_R\Bigg]\Phi'_{\uparrow}
+\frac{{\Omega}}{2}\Phi'_{\downarrow},
\\
i\frac{\partial \Phi'_{\downarrow}}{\partial t'}
=&
\Bigg[
-\frac{1}{2}\frac{\partial^{2}}{\partial x'^{2}}
-i k_{R}\frac{\partial}{\partial x'}
+\frac{k_{R}^{2}}{2}
+|\Phi'_{\downarrow}|^{2}
+g_{\uparrow\downarrow}{|\Phi'_{\uparrow}|^{2}}
\nonumber \\&-\frac{1}{\pi}
\sqrt{
|\Phi'_{\uparrow}|^{2}
+|\Phi'_{\downarrow}|^{2}
}
+ v k_R \Bigg]\Phi'_{\downarrow}
+\frac{{\Omega}}{2}\Phi'_{\uparrow}.
\end{align}
\end{subequations}
The emergence of $v$-dependent terms in transformed equations is a consequence of the breakdown of the Galilean invariance.

The ground-state solution of the transformed GP equations (\ref{tranform_gpe}a) and (\ref{tranform_gpe}b), $\phi_i'(x')$, can be
obtained using imaginary time propagation. Then, using Eq.~(\ref{phi_transf}), the moving droplet solution in the space-fixed frame at
$t=0$ is $e^{ivx} \left[\phi_i'(x')\right]_{x'=x} $. This moving droplet can be time evolved using Eqs.~(\ref{scaled_gpe}a) and (\ref{scaled_gpe}b) 
in the space-fixed frame. 
\begin{figure}
    \centering
    \includegraphics[width=\linewidth]{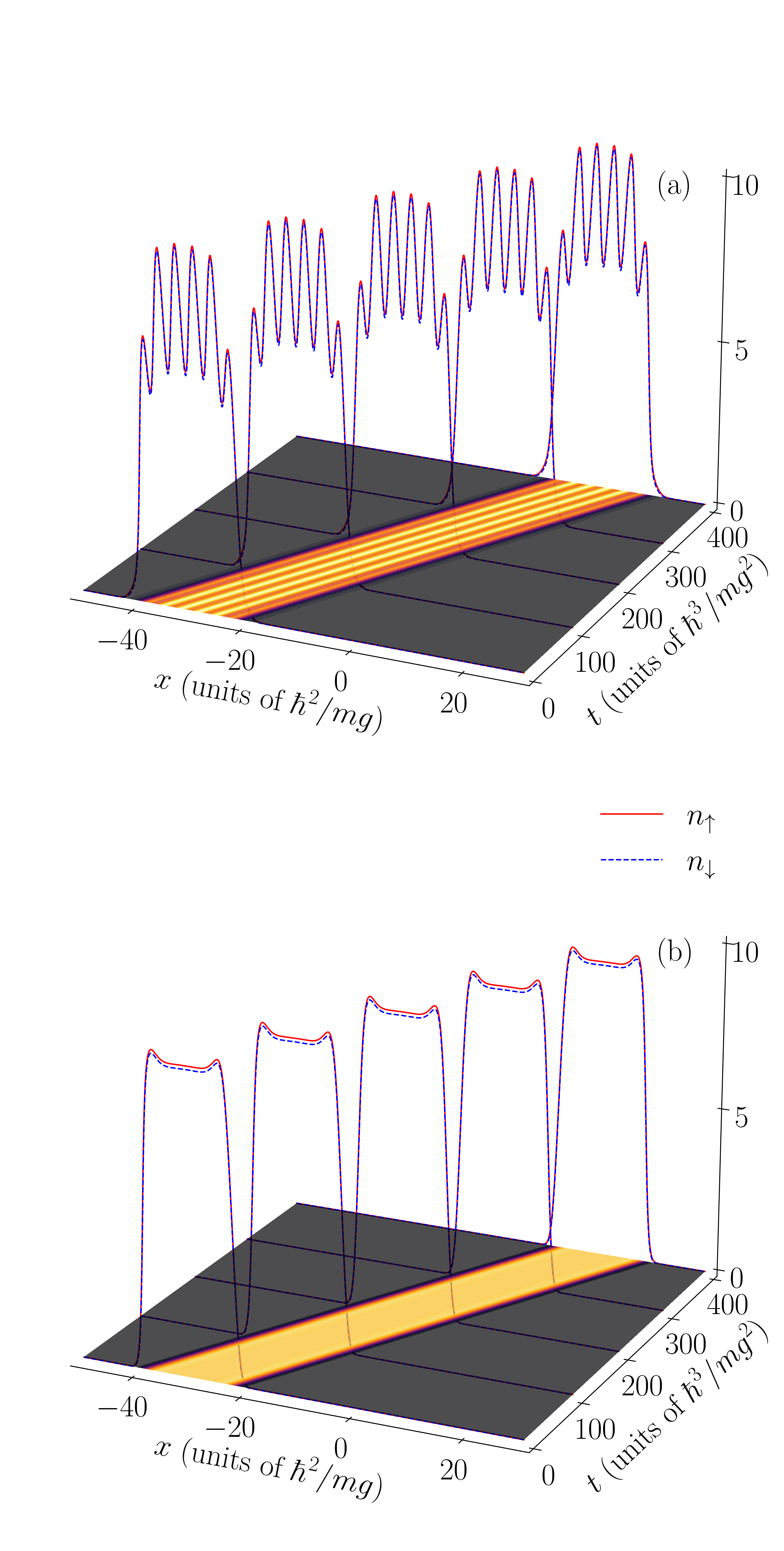}
    \caption{ Density profiles of SO-coupled droplets with $N=300$ and $g_{\uparrow \downarrow}=-0.9$ moving at velocity $\mathbf{v}=0.1 \hat{x}$ at different times for (a) a stripe droplet at $\Omega=0.5$ and (b) an unmodulated droplet at $\Omega=2$.
    The solid red and dashed blue lines represent the two component densities at successive times. The profiles translate while retaining their characteristic internal structures over the displayed time interval. The contour plots of total density are also shown in the $x$-$t$ plane.
    }
    \label{fig:mov_prof}
\end{figure}
The density profiles of the droplets traveling with ${\bf v} = 0.1 \hat{x}$ obtained using this procedure for a mixture  
at $\Omega = 0.5$ and $\Omega = 2$ are shown in Figs.~\ref{fig:mov_prof}(a) and (b), respectively. These two correspond to the two distinct ground-state phases
in the comoving frame. These profiles demonstrate that the droplets preserve their structure while traveling at a constant velocity. 

To analyze the ground states in the comoving frame, we extend the variational analysis of Sec.~\ref{Sec.III} for an infinitely extended,
SO-coupled liquid to the comoving frame. We calculate $\langle\sigma_x\rangle$, $k_1$, and $\langle \sigma_z \rangle$ as functions
of $\Omega$ to classify the various phases that can emerge in the comoving frame, where $\langle\ldots\rangle$ denotes the expectation value evaluated with respect to the ground-state solutions of Eq.~(\ref{tranform_gpe}).
Using the same  variational ansatz (\ref{par}) with $x$ replaced by $x'$, 
we obtain the variational energy per particle in the comoving frame:
\begin{eqnarray}
\label{envel}
    \mathcal{E}(n_0,k_1,\beta)&\simeq&
    \frac{k_R^2}{2}-\frac{f(\beta,n_0) k_1^2}{2k_R^2}+(2 \beta +1) G_1-\frac{\Omega}{2k_R}\\&\times&\sqrt{k_R^2-k_1^2}-\frac{\sqrt{n_0}}{2\pi}\left(\frac{4}{3}+\beta\right) -v k_1 \sqrt{1-4\beta}, \nonumber
\end{eqnarray}
where the last term is the contribution to the variational energy arising from the velocity-dependent term;  $G_1=n_0(1+g_{\uparrow\downarrow})/4$; $G_2=n_0(1-g_{\uparrow\downarrow})/4$; $\beta = |c_1|^2|c_2|^2$; and $f(\beta,n_0)$ is
defined in Eq.~(\ref{f_beta}). $\mathcal{E}(n_0,k_1,\beta)$ can be numerically minimized to obtain the average saturation density $n_0$, $k_1$, and $\beta$. These values can then be used to calculate $\langle \sigma_z\rangle$, $\langle \sigma_x\rangle$  via Eqs.~(\ref{sigmaz}) and (\ref{sigmax}).
\begin{figure}
    \centering
    \includegraphics[width=\linewidth]{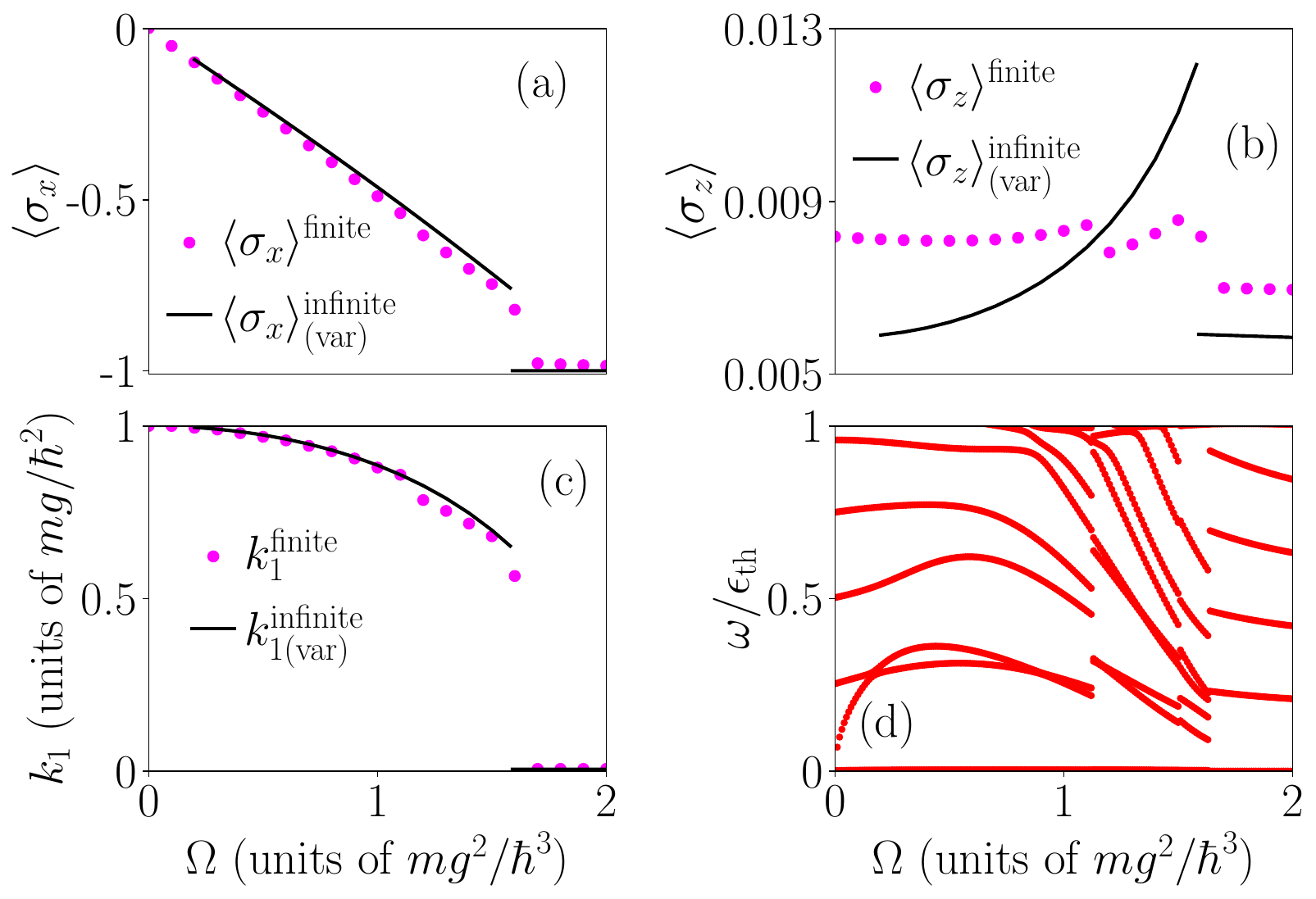}
    \caption{ Ground-state properties as a function of the Rabi coupling $\Omega$ at a fixed velocity $v=0.1$ in the comoving frame: (a) the transverse polarization $\langle \sigma_x \rangle$, (b) the velocity-induced longitudinal polarization $\langle \sigma_z \rangle$, and (c) the ground-state momentum $k_{1}$. In (a)–(c), solid magenta circles denote the values for the ground-state solutions obtained by numerically solving the GP Eq.~(\ref{tranform_gpe}) for $N=300$ and $g_{\uparrow \downarrow} =-0.9$, while black line represent results from minimizing the variational energy (\ref{envel}). (d) The low-lying excitation spectrum of the finite-size droplet for $N = 300$ and $g_{\uparrow \downarrow} =-0.9$ in the comoving frame, plotted as a function of $\Omega$ in units of the particle emission threshold.
    }
    \label{fig:ex_vel_omega}
\end{figure}
These variational estimates are shown by black lines in Figs.~\ref{fig:ex_vel_omega}(a)--(c). Unlike static droplets, moving droplets feature a finite (bare) spin polarization $\langle \sigma_z\rangle$ (see Fig.~\ref{fig:ex_vel_omega}(b)). Furthermore, $\langle\sigma_x\rangle$, $\langle \sigma_z\rangle$, and $k_1$ change discontinuously across the critical point of transition ($\Omega^{\rm var}_{\rm{cr}}=1.59$) from the
ST to the PW droplet. In the PW phase as only the global minimum of the lowest dispersion branch is occupied, $k_1 = \langle \sigma_z\rangle$. For the finite-size droplets, these expectations in the comoving frame at $v=0.1$ are also plotted in Figs.~\ref{fig:ex_vel_omega}(a)--(c) (see solid magenta circles).
The number of peaks changes as  $\Omega$ is varied, which results in the discontinuities observed in Fig.~\ref{fig:ex_vel_omega}(b) below $\Omega_{\rm cr}^{\rm num} = 1.64$.

We now analyze the stability of the moving droplets by calculating the excitation spectrum in the comoving frame.
To this end, we linearize Eq.~(\ref{tranform_gpe}) by considering fluctuations 
around the ground state $\phi_{i}'(x')$ using 
\begin{equation}
    \Phi_i'(x',t') = e^{-i \mu' t'}\Big[\phi_i'(x') + \delta{\phi'}_{i}(x',t')\Big],
\end{equation}
where $\mu'$ is the chemical potential in the comoving frame. We then use the Bogoliubov transformation to obtain the BdG equations in
the comoving frame. These BdG equations can be obtained from Eq.~(\ref{bdg}) by changing $x$ to $x'$ and with $\mathcal{ M}$ and $\mathcal{N}$
now defined as
\begin{subequations}
    \begin{align}
        \mathcal{M} &= -\frac{1}{2}\frac{\partial^2}{\partial x'^2}+i k_{R}\frac{\partial}{\partial x'}+ \frac{k_{R}^2}{2}+  2|\phi_\uparrow'|^2+g_{\uparrow\downarrow}|\phi_\downarrow'|^2 \nonumber \\& \qquad  -\mu' -\frac{1}{2\pi \sqrt{n}}\left( 3|\phi_\uparrow'|^2 +2|\phi_\downarrow'|^2\right) -v k_R,\\ 
        \mathcal{N} &= -\frac{1}{2}\frac{\partial^2}{\partial x'^2}  -i k_{R}\frac{\partial}{\partial x'}+ \frac{k_{R}^2}{2}+ 2|\phi_\downarrow'|^2+g_{\downarrow\uparrow}|\phi_\uparrow'|^2 \nonumber \\& \qquad -\mu' -\frac{1}{2\pi \sqrt{n}} \left(3|\phi_\downarrow'|^2 +2|\phi_\uparrow'|^2\right)+v k_R.
    \end{align}
\label{matrix_element}
\end{subequations}
The particle emission threshold for an SO-coupled droplet in the comoving frame is $\epsilon_{\rm th}=\epsilon_{\rm min}-\mu' $, with $\epsilon_{\rm min}$ now
defined as the global minimum of the lower single-particle dispersion branch
\begin{equation}
\epsilon_{-} = \frac{k'^2+k_R^2}{2} -\sqrt{k_R^2(k'+v)^2+\left(\frac{\Omega}{2}\right)^2} \label{ethreshold}.
\end{equation}
The excitation spectrum evaluated by numerically solving the BdG equation in the comoving frame for $v =0.1$
as a function of $\Omega$ is shown in Fig.~\ref{fig:ex_vel_omega}(d). We find all the excitation energies are real with  $\omega_l \int dx'  \sum_{i{=\uparrow,\downarrow}}\left(|u_i^l|^2-|v_i^l|^2\right)>0$ for all $\omega_l\ne0$, thus confirming the energetic and dynamical stability of the moving droplets. Again, the spectrum bears
the signature of a change in the number of peaks from six for $0<\Omega \le 1.12$, to five for $1.13\le\Omega\le 1.5$, and then
to four for $1.51\le\Omega\le1.63$ in the traveling supersolid stripe droplet in the form of discontinuities in the spectrum. 


\begin{figure}[!ht]
    \centering
    \includegraphics[width=\linewidth, height=10cm, keepaspectratio]{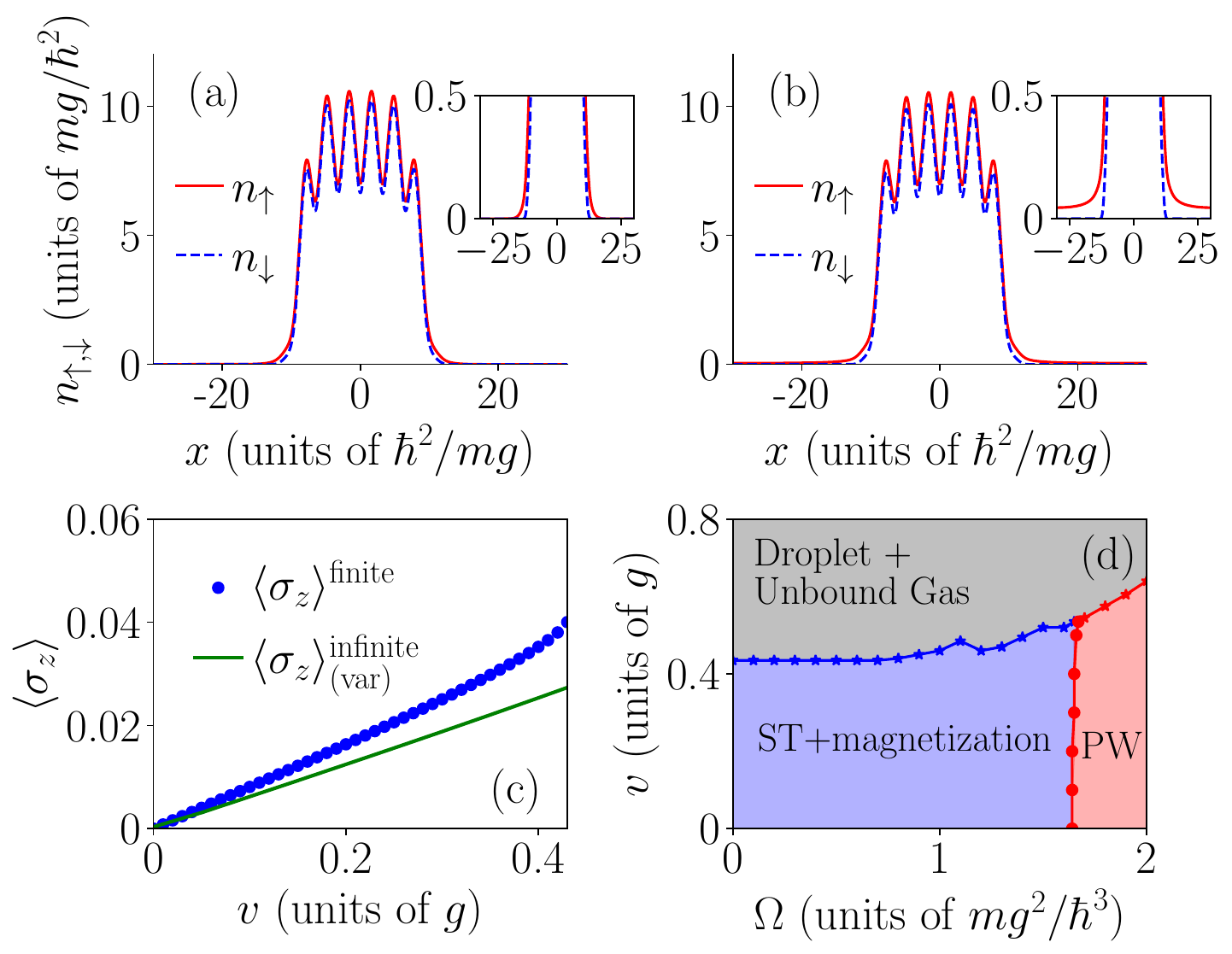}
    \caption{ Component density profiles, spin polarization, and ground-state phase diagram in the comoving frame. Component density profiles are shown for a particle number $N=300$, inter-component interaction $g_{\uparrow \downarrow}=-0.9$, and Rabi coupling $\Omega=0.5$ at velocities (a) $v=0.35$ and (b) \(v=0.44\). The insets provide magnified views illustrating the detailed spatial behavior of the densities. (c) Spin polarization $\langle \sigma_z \rangle$ as a function of $v$ for $\Omega=0.5$ and $g_{\uparrow\downarrow}=-0.9$. The solid green line is obtained from minimizing the variational energy functional (\ref{envel}), which is independent of particle number, and is compared against the polarization of the ground-state solutions of GP Eq.~(\ref{tranform_gpe}) for $N=300$. (d) Ground-state phase diagram in the $(\Omega,v)$ plane for $N=300$ and $g_{\uparrow\downarrow}=-0.9$.
    The blue region corresponds to magnetized ST droplets, the upper region represents a droplet core immersed in an unbound gas of the majority component, and the red region denotes an unmodulated PW droplet.
    }
    \label{unbound_dens}
\end{figure}

\begin{figure}[!ht]
    \centering
    \includegraphics[width=\linewidth]{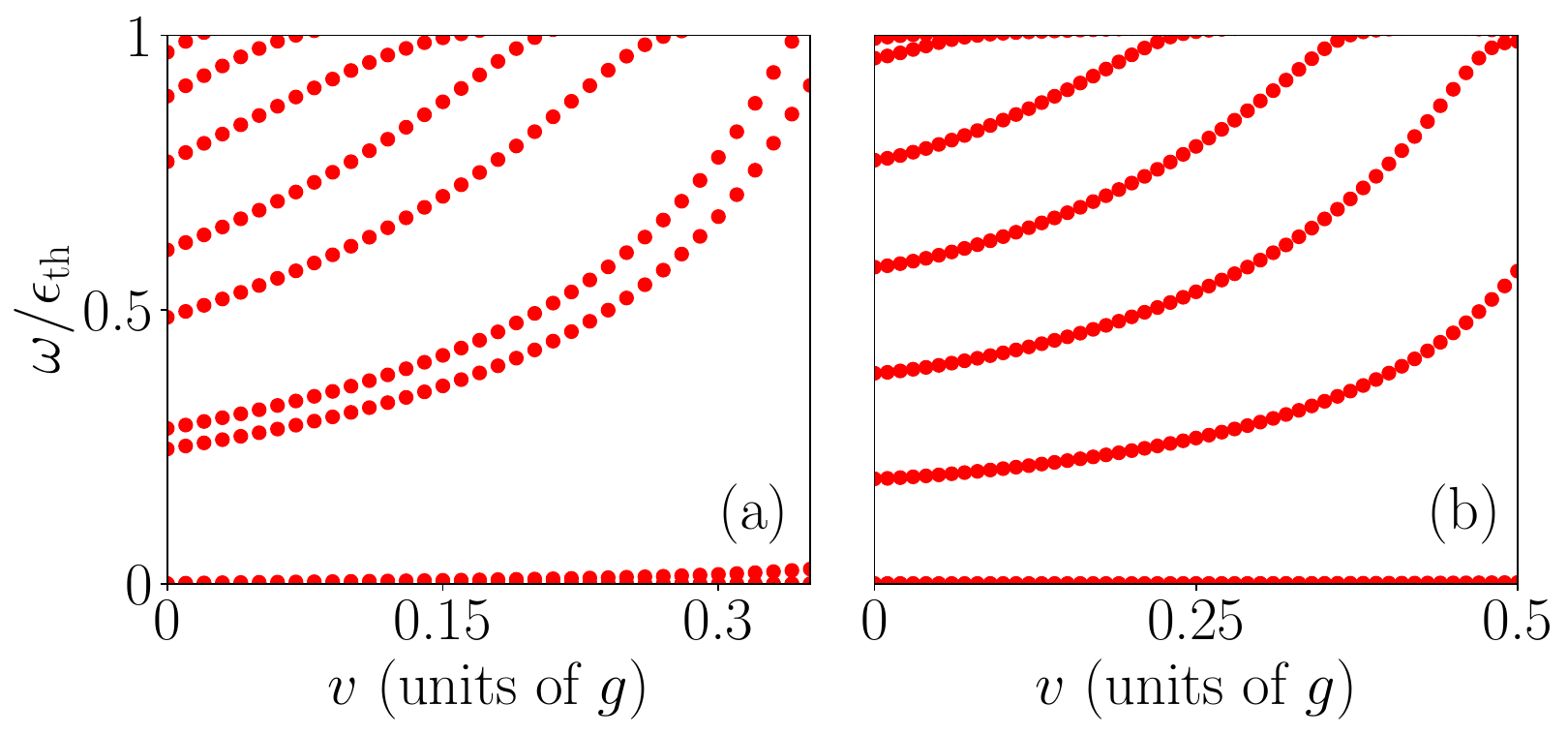}
    \caption{ Excitation spectrum as a function of velocity $v$ for fixed parameters $N=300$ and $g_{\uparrow \downarrow}=-0.9$ in the (a) ST droplet phase at $\Omega=0.5$ and (b) PW droplet phase at $\Omega=2$.
    }
    \label{fig:ex_vel}
\end{figure}

 As $v$ increases, the system remains bound up to a critical velocity ($v_{\rm{cr}}$), above which the excess atoms of the majority component become unbound.
Consequently, for $v>v_{\rm{cr}}$, the ground-state solution $\phi_i'(x')$ consists of a periphery of unbound atoms surrounding a large number of bound atoms within a droplet core. For instance, for the finite-size droplet at $\Omega = 0.5$, the critical velocity is $v_{\rm cr} = 0.43$. 
This behavior is illustrated in Figs.~\ref{unbound_dens}(a) and \ref{unbound_dens}(b) by two representative density profiles of the ST droplet traveling in the bound and unbound regimes, respectively. Above the critical velocity, the unbound atoms tend to occupy the entire simulation domain (see inset of Fig.~\ref{unbound_dens}(b)). 
The magnetization increases with increasing velocity. The magnetization values for an infinitely extended liquid, obtained by minimizing the variational energy (\ref{envel}), and for the finite-size 
droplet as a function of $v$ are shown in Fig.~\ref{unbound_dens}(c). The particle emission threshold (\ref{ethreshold}) decreases and approaches zero as velocity increases (not shown here). 
The ground-state phase diagram in $\Omega$-$v$ plane, obtained using imaginary-time propagation, is shown in Fig.~\ref{unbound_dens}(d) for $0 \leq \Omega \leq 2$.
We observe that as $\Omega$ increases, the critical velocity increases, except at critical $\Omega$ values across which the number of stripes decreases by one
in the ST droplet phase. We also calculated the excitation spectrum as a function of velocity $v$ for $\Omega=0.5$ (ST droplet phase) and $\Omega=2$ (PW droplet phase) in Figs.~\ref{fig:ex_vel}(a) and \ref{fig:ex_vel}(b), respectively.
As the velocity increases, the excitation frequencies in units of particle emission threshold shift upward, and the number of excitations below the particle emission threshold decreases. 
We calculated the excitation spectrum up to the velocities where we reliably captured the zero-energy modes in the excitation spectrum for spatial step size $\Delta x=0.02$. As one approaches $v_{\rm{cr}}$, $\Delta x$ must be decreased further to resolve the excitation spectrum more accurately.

\begin{figure}[!ht]
    \centering
   \includegraphics[width=\linewidth]{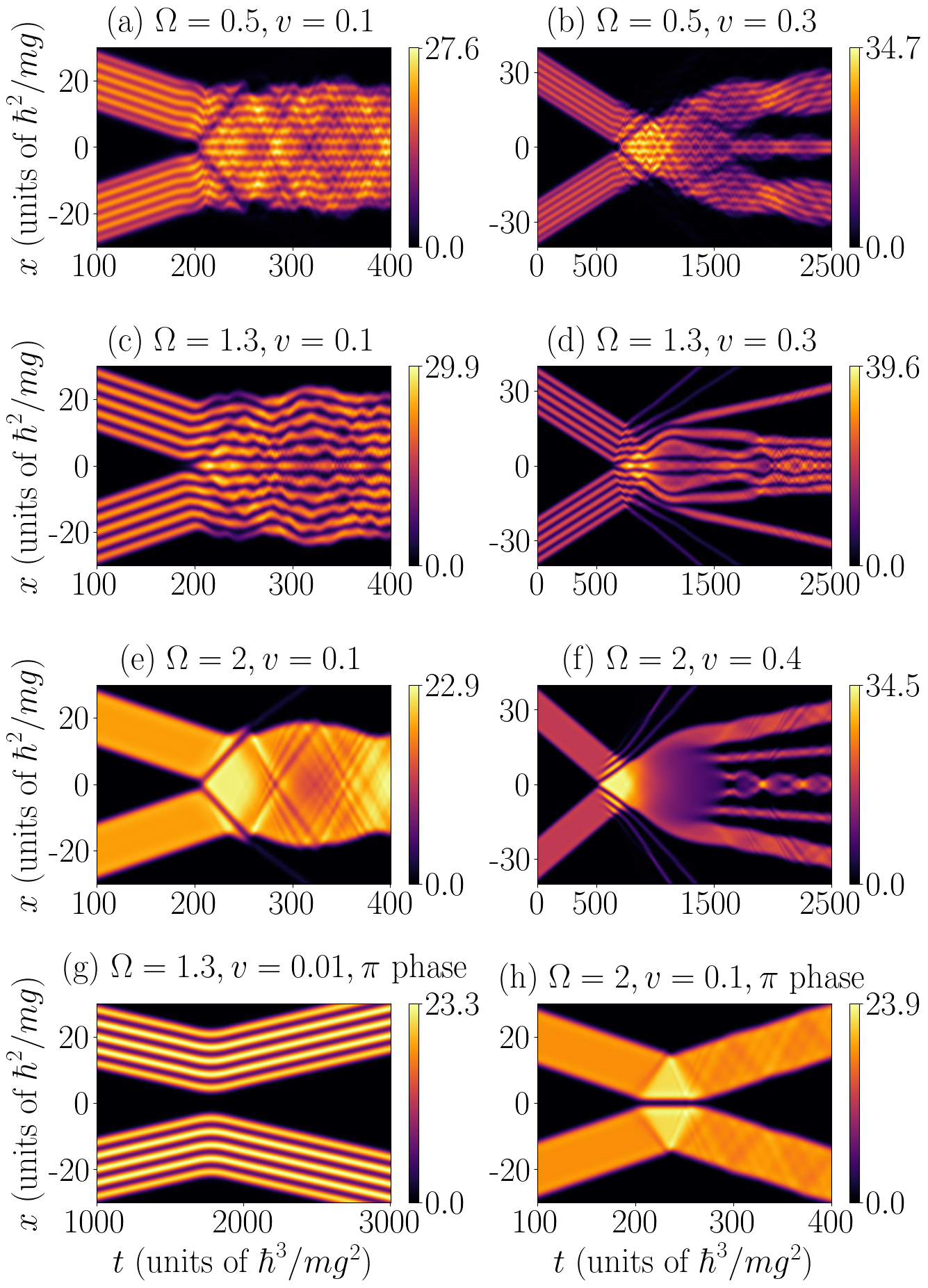}
    \caption{ Time evolution of the total density profiles for two droplets moving with equal and opposite velocities. 
    For each droplet, $N = 300$ and $g_{\uparrow\downarrow} = -0.9$.
    The two droplets are in-phase in (a)-(f) and have a phase difference of $\pi$ in (g) and (h).
    }
    \label{collision_droplet}
\end{figure}
Lastly, we also studied head-on collisions between pairs of both in-phase and out-of-phase droplets.
Collisions between two in-phase droplets can result in merging, fragmentation, or both. Figs.~\ref{collision_droplet}(a)--\ref{collision_droplet}(f) show the time evolution of the density profiles for in-phase pairs of colliding droplets at different velocities and coherent coupling strengths, with $N=300$ and $g_{\uparrow \downarrow}=-0.9$. The slow-moving droplets with $v=0.1$ shown in Figs.~\ref{collision_droplet}(a), \ref{collision_droplet}(c), and \ref{collision_droplet}(e) for $\Omega=0.5$, $\Omega=1.3$, and $\Omega=2$, respectively, collide to primarily form a single static droplet in an excited state. At relatively higher velocities, the collisions lead to fragmentation, as shown in Figs.~\ref{collision_droplet}(b) ($\Omega = 0.5$, $v= 0.3$), \ref{collision_droplet}(d) ($\Omega= 1.3$, $v =0.3$), and \ref{collision_droplet}(f) ($\Omega= 2, v= 0.4$). The majority of the atoms are carried away by the mobile droplet fragments in Figs.~\ref{collision_droplet}(b) and \ref{collision_droplet}(f), with only a minority remaining in the static central fragment. 
In contrast to the inelastic collisions of in-phase droplets, the out-of-phase droplets at very small velocities can rebound after colliding, keeping their structures intact as shown in Fig.~\ref{collision_droplet}(g) ($\Omega= 1.3$, $v =0.01$) for ST and Fig.~\ref{collision_droplet}(h) ($\Omega= 2$, $v =0.1$) for PW droplets.
In in-phase and out-of-phase collisions, the total density after the collision at $x=0$ shows a node (local density minimum) 
or an antinode (local density maximum) depending on the number of stripes in the traveling ST droplets. 
As shown in Figs.~\ref{collision_droplet}(a) and (b), six stripes result in a node post-collision, while five stripes in Figs.~\ref{collision_droplet}(c)-\ref{collision_droplet}(d) lead to an antinode. 
Additionally, the PW traveling droplets exhibit an antinode in in-phase collisions and a node in out-of-phase collisions, as is seen in Figs.~\ref{collision_droplet}(e) and \ref{collision_droplet}(h), respectively.

\section{Summary and Conclusions}\label{VI}
In conclusion, we have investigated the static and traveling droplet phases and their collective excitations in a 1D spin-orbit-coupled Bose-Bose mixture with repulsive 
intra- and attractive inter-component coupling strengths near the mean-field collapse threshold. We used a variational approach to calculate the ground-state phase diagram of the
mixture in the thermodynamic limit. For the parameters considered, the system undergoes a direct transition from a stripe droplet to a zero-momentum droplet with increasing Rabi 
coupling strength. The variational estimates of the transition point, momentum, and spin polarization agree well with the numerical results.

For finite-size droplets, we find that the stripe regime supports various stationary states with different numbers of stripes at a fixed Rabi coupling strength. The 
corresponding spectra show that these higher-energy states can remain locally metastable. Importantly, discontinuities are observed in the ground-state excitation spectrum when 
the number of stripes in the ground state changes as either the Rabi coupling or the number of atoms is varied. In contrast, the excitation spectrum of a stationary state 
with a fixed number of stripes evolves continuously.

We further examined droplets with finite velocities using the Galilean-boosted framework. In the comoving frame, two finite
magnetization stable ground-state droplet phases emerge, namely stripe and plane-wave. These droplets remain bound only below
a critical speed, above which atoms of the majority component leak out of the droplet. Finally, head-on collisions between two in-phase droplets exhibit rich merging and splitting dynamics. 

\begin{acknowledgments}
K.S. acknowledges support from the Science and Engineering Research Board, Department of Science and Technology, Government of India through Project No. SRG/2023/001569.
\end{acknowledgments}

\bibliography{ref}{}
\bibliographystyle{apsrev4-2}
\end{document}